\documentclass[preprint,aps,amsmath,amssymb]{revtex4}

\usepackage{graphicx}
\usepackage{dcolumn}
\usepackage{bm}

\begin{document}

\title{
Simulation for the oblique impact of a lattice system
}

\author{
Hiroto \textsc{Kuninaka}
\footnote{E-mail address: kuninaka@yuragi.jinkan.kyoto-u.ac.jp} and
Hisao \textsc{Hayakawa}
\footnote{E-mail address: hisao@yuragi.jinkan.kyoto-u.ac.jp}
}

\address{%
GSHES, Kyoto University, Sakyo-ku, Kyoto 606-8502
}

\date{\today}
\begin{abstract}
The oblique collision between an elastic disk and an elastic
 wall is numerically studied.
 We investigate the dependency of the tangential coefficient
 of restitution on the incident angle of impact.
 From the results of simulation,
 our model reproduces experimental results
 and can be explained by a phenomenological theory
 of the oblique impact.
\end{abstract}
\maketitle
\newpage
\section{Introduction}
Collisions are common phenomena in nature.
 In macroscopic scales,
 we often see collisions of balls in sports such as the baseball
 and the billiard. 
 In such collisions, the initial kinetic energy of material dissipates  
 into internal degrees of freedom.
 A part of total energy is distributed into the
 translational motion and the rotational motion, while
 the other part is dissipated as elastic vibration,
 sound emission, heat, etc.
 As a result, collisions of macroscopic 
 material are always inelastic.

Inelastic collisions play important roles
 in granular materials\cite{granules,degennes}.
 Characteristic behaviors of granular material come from
 inelastic collisions among particles.  
 The {\it Distinct Element Method} (DEM) is
 a standard method of simulation for the granular materials\cite{dem}.
 DEM contains some phenomenological parameters such as 
 the Coulomb's coefficient of friction, dashpots, and so on.
 Nobody can determine such the viscoelastic parameter
 from the first principle.
 Even the determination of the simplest parameter,  
 the coefficient of restitution (COR), is difficult.

\begin{figure}[thbp]
  \begin{center}
   \includegraphics[width=1.0\textwidth]{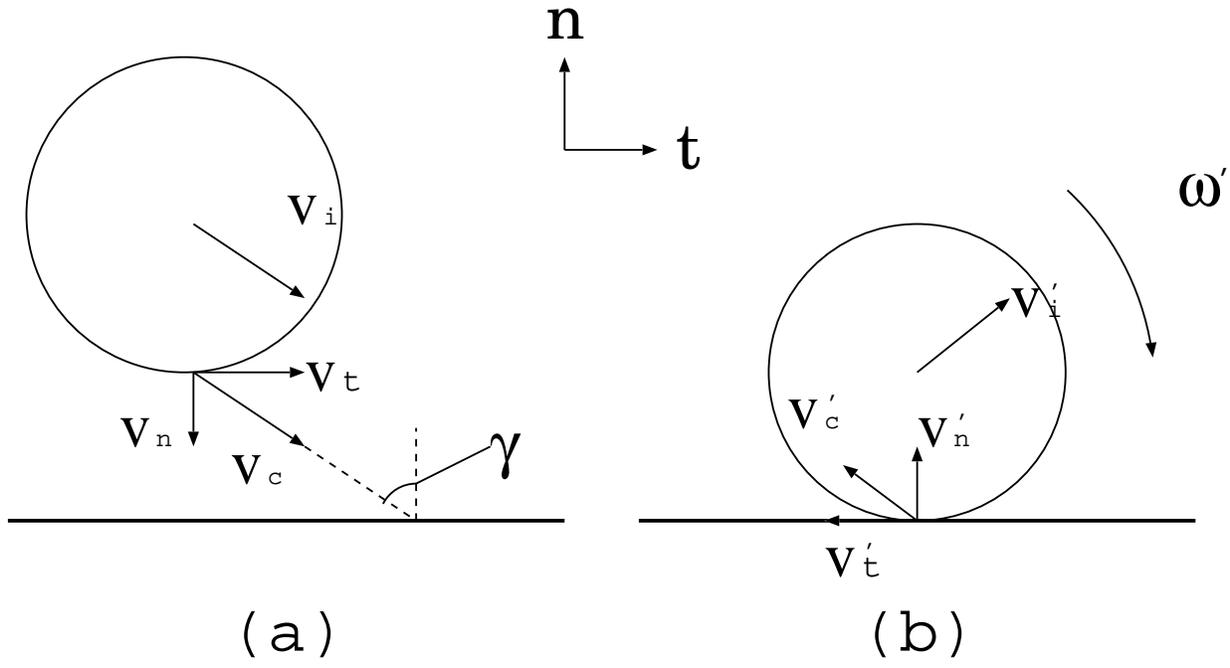}
   \caption{H. Kuninaka and H. Hayakawa}
   \label{figure0}
  \end{center}
\end{figure}

 COR is a familiar parameter
 which is introduced in text books of the elementary physics. 
 The normal COR is defined by
 the ratio of the normal components of the collision velocity
 before and after collision.
 Figure \ref{figure0} is the schematic figure
 that a sphere is colliding
 with a stationary wall with initial velocity
 of its center of mass, ${\bf v}$.
 The prime denotes post-colliding quantities.
 The coefficient of normal restitution $e$ is defined by
\begin{equation}
{\bf v_c^{'}} \cdot {\bf n}=-e {\bf v_c} \cdot {\bf n},
\end{equation}
 where ${\bf v_c}$ and ${\bf v_c^{'}}$ are respectively the velocity
 of the contact point before and after the collision.
 $e$ is assumed to be $0 \le e \le 1$.
 Historically, COR was first introduced by Newton\cite{newton}.
 Though many text books of elementary physics state
 that $e$ is a material constant,
 many experiments and simulations show that $e$ decreases
 as the impact velocity increases\cite{johnson,goldsmith,stronge
,sonder,bridges,supulver,kuwabara}
 except for  the idealized situation of a collision between two rods,
 in which $e$ may be determined by the ratio of
 length of the colliding rods\cite{giese,aspel,basile,sugiyama,nagahiro}.
 Recently, Louge and Adams have reported
 that $e$ can exceeds unity in an oblique impact\cite{louge02}.
 The mechanism of the exotic behavior has not been clarified yet.
 Thus, study of collisions is not matured and has room for development
 though the subject itself is familiar even for high school students.

In addition to $e$, the tangential COR $\beta$
 is also important to characterize oblique impacts,
 where $\beta$ is defined as
\begin{equation}
{\bf v_c^{'}} \cdot {\bf t}=-\beta {\bf v_c} \cdot {\bf t},
\end{equation}
 where ${\bf v_c^{'}}$ and ${\bf t}$ are
 the post-collisional velocity at
 the contact point after collision and
 the unit tangential vector, respectively. 
 $\beta$ is a function of the angle of incidence
 $\gamma$ which is defined as $\gamma=\arctan(v_t/v_n)$
 with $v_n={\bf v}_c \cdot {\bf n}$ and $v_t={\bf v}_c \cdot {\bf t}$
 and is believed
 that possible values of $\beta$ lie
 between -1 and 1\cite{walton,labous,foerster,lorentz,gorham}.

 The aim of our study is to investigate $\beta$ in detail
 from numerical simulation.
 We study the relation between $\beta$
 and the angle of incidence in oblique collision in this paper.
 The organization of this paper is as follows.
 In \S 2, we will review
 the current status of the researches about inelastic collisions.
 In \S 3, we introduce our numerical model 
 and setup of the simulation.
 Section 4 is the main part of this paper
 where we summarize the results of our simulation
 and compare them with the theoretical outcome.
 Section 5 is devoted to the discussion of our results.
 The last section is the conclusion remarks.
 In Appendix \ref{appendixA}, we will give a description of
 the theory of the oblique impact.
 In Appendix \ref{appendixB}, the derivation of Poisson's ratio
 of the square lattice system will be introduced.
\section{Review}
In this section, we review current status of the study
 of inelastic collision.
 At the end of the 19th century,
 it became clear experimentally that COR depends on the
 relative velocity of the colliding materials\cite{vincent}.
 In the normal collision of spheres, it is believed that
 COR obeys $1-e \varpropto v^{1/5}$ when
 impact velocity $v$ is low enough while $e \varpropto v^{-1/4}$
 when $v$ exceeds
 the critical value of plastic deformations.
 In fact, Sondergaard {\it et al.}\cite{sonder} performed
 an impact experiment of ball bearings and glass spheres
 on lucite or aluminum plates and confirmed
 the dependency on the impact velocity as $e \varpropto v^{-1/4}$
 in the high speed impact.
 Johnson explained this dependency
 using the dimensional analysis\cite{johnson}.
 His analysis was consistent with the experimental data.
 Kuwabara and Kono\cite{kuwabara} investigated COR for low speed impact
 and derived the theoretical expression $1-e \varpropto v^{1/5}$
 which is consistent with their experimental results.
 Recently, some papers about quasi-static theory are published
 which are consistent
 with Kuwabara and Kono\cite{morgado,brilliantov96,schwager,ramirez}. 

 On the other hand, Gerl and Zippelius performed
 simulation of a two-dimensional collision
 of an elastic disk with an elastic wall\cite{gerl}.
 The present authors  performed  
 two-dimensional simulations
 and confirmed that elastic models including that of Gerl and Zippelius
 is not appropriate
 to characterize the quasi-static region\cite{kuninaka,kuninaka2,hisao}.
 
To characterize the oblique collision,
 Walton introduced three parameters:
 the coefficient of normal restitution $e$, 
 the coefficient of Coulomb's friction $\mu_0$,
 and the maximum value of the coefficient of tangential
 restitution $\beta_0$\cite{walton}.
 Experiments have supported that his characterization
 adequately capture the essence of binary collision of spheres or
 collision of a sphere on a flat plate\cite{labous,foerster,lorentz,gorham}.
 Walton derives 
\begin{equation}\label{walton}  
\beta \simeq
\left\{
 \begin{array}{ll}
  -1+\mu_0 (1+e) \cot\gamma \left(1+\frac{mR^2}{I}\right)
&  (\gamma \ge \gamma_{0})\\
   \beta_{0} &  (\gamma \le \gamma_{0}),
 \end{array}
\right.
\end{equation}
where $\gamma_0$ is the critical angle, and $m$, $R$, and $I$ are
 mass, radius and moment of inertia of spheres respectively.\cite{walton}
 Experimental results are consistent
 with eq.(\ref{walton})\cite{labous,foerster,lorentz,gorham}.
 Meanwhile, Maw {\it et al.} extended
 the Hertz theory of impact\cite{landau,love,hills,hertz}
 and established the theory of the oblique impact to be consistent
 with their experimental results\cite{mbf76,mbf81}.
 The theory by Maw {\it et al.} has several advantages:
 (i) The physical mechanism is included in their theory but not
 in Walton's argument.
 (ii) Anomalous behavior near $\gamma=0$ can be described
 in their theory.
 The disadvantage of their argument is that we cannot summarize
 the result of theory in a concise form as in eq.(\ref{walton}).
 Thus, we still use both Walton's expression
 and the theory by Maw {\it et al.}.
 The details of the theory by Maw {\it et al.}
 and its application to our results is summarized
 in Appendix \ref{appendixA}.
\section{Our Models}
In this section, let us introduce our numerical model.
 Our numerical model consists of an elastic disk and an elastic wall
(Fig. \ref{figure1}).
 Both of them are composed of randomly distributed 1600 mass points.
 We use the Delaunay triangulation algorithm
 to connect all mass points with nonlinear springs\cite{delaunay}.
 The spring interaction between connected mass points
 is described as
\begin{equation}\label{interaction}
 V(x)=\frac{1}{2} k_{a} x^{2}+\frac{1}{4} k_{b} x^{4},
\end{equation}
 where $x$ is a stretch from the natural length of spring,
 and $k_{a}$ and $k_{b}$ are the spring constants.
 In most of simulations, we adopt $k_a = 1.0 \times m c^2/R^2$
 and $k_b = 1.0 \times 10^{-3} m c^2/R^4$, respectively.
 The width of the wall is 4 times as long as the diameter of the disk
 while the height of the wall is same as the diameter of the disk.
 Two sides of the wall are fixed.

 The interaction between the disk and the wall during a collision
 is introduced as follows. Figure \ref{figure2} is the schematic
 figure of the interaction of surface mass points
 of the disk and the wall.
 When the distance $l$ between the lower edge of the disk 
 and the surface of the wall
 becomes less than the cutoff length
 which is the mean value
 of natural lengths of all springs,
 the surface particles of the disk
 feel the repulsive force,
 ${\bf F}(l)=aV_0\exp(-al){\bf n}$,
 where $a$ is $300/R$, $V_0$ is
 $amc^2R/2$, $m$ is the mass of the particle,
 $R$ is the radius of the disk, 
 $c=\sqrt{E/\rho}$,
 $E$ is Young's modulus, and $\rho$ is the density, 
 ${\bf n}$ is the normal unit vector to the surface.
 The reaction forces applied to the two points of 
 the surface of the wall (point 1 and 2) are decided by the balance
 of the torques as 
 ${\bf F_1}(l)=-F(l){\bf n}/(1+l_1/l_2)$
 and ${\bf F_2}(l)=-F(l){\bf n}/(1+l_2/l_1)$,
 where $l_i(i=1,2)$ is the distance
 between the point $p$ and the point $i$ (see Fig. \ref{figure2}).

In this model, roughness of the surfaces is important
 to make the disk rotate after collisions.
 To make roughness, at first,
 we generate normal random numbers whose average value is $0$ and then
 make the initial position of particles on surfaces of
 both the disk and the wall deviate with them.
 We choose the standard deviation of the normal random numbers
 $\delta$ as $\delta = 3 \times 10^{-2} R$,
 where $R$ is the radius of the disk.
 All of the data presented here are obtained from the average 
 of 100 samples in random numbers.

For random lattice model,
 it is impossible to determine Poisson's ratio $\nu$ and Young's modulus $E$
 theoretically.
 When we determine Poisson's ratio $\nu$ and Young's modulus $E$ of this model,
 we add the viscous force term
 in equation of motion which is proportional to the relative velocity
 of two connected mass points. 
 By stretching the band of random lattice
 and calculating the ratio of the strains
 in the vertical and horizontal directions to the force
 and the ratio of the strain to the force
 when the vibration stops,
 we can obtain Poisson's ratio and Young's modulus.
 Figure \ref{ranst}(a) and \ref{ranst}(b)
 are snap shots of the bands of random lattice made of 348 mass points
 before and after adding the force $F=3.0 \times 10^{2}mc^{2}/R$, respectively.
 We change the force from $2.0 \times 10^{2} mc^{2}/R$
 to $3.0 \times 10^{2} mc^{2}/R$
 and average 10 samples of results
 to obtain $\nu=(7.50 \pm 0.11) \times 10^{-2}$
 and $E=(9.54 \pm 0.231)\times 10^{3}mc^{2}/R^{2}$,
 respectively.

For comparison, we introduce other two lattice models for elastic disks:
 triangular lattice and square lattice disk(Fig.\ref{lattice}).
 To investigate the effect of the structure of the disk,
 the wall is same as random lattice model.
 In order to remove anisotropies of lattice structure, 
 we put an exterior layer which is same as random lattice disk
 around the disk and bind all of them
 using the Delaunay triangulation algorithm.
 The triangular lattice disk is made by replacing
 the internal structure of the random lattice disk
 with the triangular lattice.
 Total number of mass points is same as that of random disk.
 Poisson's  ratio and Young's modulus of the triangular lattice
 can be calculated theoretically
 as $1/3$ and $2 k_a/\sqrt{3}$ 
 in the continuum limit respectively\cite{hoover}. 
 The square lattice disk is made by replacing the internal 
 structure of the random lattice disk
 with the square lattice and connecting all the mass points by
 the Delaunay triangulation algorithm. We introduce two spring constants:
 $k_a=k_1$ for nearest neighbor interaction and $k_a=k_2$
 for next-nearest neighbor interaction.
 In the continuum limit,  Young's modulus $E$\cite{yhayakawa}
 and Poisson's ratio $\nu$ (see Appendix \ref{appendixB})
 of the square lattice are expressed as
\begin{eqnarray}
\frac{1}{E}=\frac{k_{1}+k_{2}}{k_{1}(k_{1} + 2 k_{2})}
+\frac{k_{1}-2k_{2}}{k_{1}k_{2}}n_{x}^{2} n_{y}^{2},\label{young} \\
\nu = \frac{k_{2}^2 + (k_{1}^2-4k_{2}^2)n_{x}^2 n_{y}^2}
{k_{2}(k_{1}+k_{2})+(k_{1}^2-4 k_{2}^2) n_{x}^2 n_{y}^2}\label{poisson},
\end{eqnarray}
 where $n_x$ and $n_y$ are the unit normal vectors horizontal and vertical
 to the collisional plane.
 The derivation of eqs.(\ref{young}) and (\ref{poisson}) is 
 presented in Appendix \ref{appendixB}.
 To recover the orientational symmetry and introduce roughness
 on the surface of disks, we introduce one-layer random lattice
 on the surface of the disks.

We scale the equation of motion for each particle
 using the radius of the disk $R$ as the scale of length
 and the velocity of elastic wave $c=\sqrt{E/\rho}$
 as the scaling unit of velocity.
 As the numerical scheme of the integration,
 we use the fourth order symplectic numerical method
 with the time step $\Delta t \simeq 10^{-3}R/c$.
\section{Results}
 In this section,
 we explain the results of our simulation.
 The angle of incidence $\gamma$ is ranged from
 $5.7^{\circ}$ to $80.5^{\circ}$
 while the normal component of velocity is fixed as $0.1c$.
 The colliding disk has no internal vibration and rotation
 at release time.
 In order to eliminate the effect of the initial configuration
 of mass points, we prepare 100 samples of disk as the initial condition
 by using 100 sets of normal random numbers and average data of all samples.

Figure \ref{figure3} shows the relation between $\cot \gamma$
 and the coefficient of  tangential restitution $\beta$.
 In this figure,
 cross points are the result of the impact
 between random lattice disk and wall,
 and broken lines are eq.(\ref{walton}).
 In eq.(\ref{walton}) we use the value $e=0.8$
 which is the approximate mean value of $e$
 in the range $2.5 \leq \cot \gamma \leq 6$ in Fig.\ref{figure4}.
 The result of simulation shows
 that $\beta_0$ is $0.56$
 and $\mu_0$ is $0.18$
 which are close to the values observed in experiments
 of three dimensional impacts\cite{labous,foerster}.
 Thus, we reproduce experimental tendencies
 of the oblique collision
 with the random lattice model\cite{labous,foerster}.
 Stars are the results of random lattice disk
 without roughness on the surface,
 in which $\beta$ is close to $-1$.
 From this result, one can see that roughness on the surface is important 
 for the rotation of the disk after collision.  
 Plus points in Fig.\ref{figure3} are the result
 of the triangular lattice model
 where the orientation of initial disks is same as that
 in Fig.\ref{lattice}(a).
 In this model, $\beta$ takes negative values in all range
 of the angle of incidence.
 This means that the disk made of triangular lattice is easy to slip
 on the surface.
 In addition, the result strongly depends on the initial orientation
 of the disk. 
 Thus, the model of triangular lattice is inadequate
 to reproduce the tendency of experimental data.

Figure \ref{figure4} shows the relation between $\cot \gamma$ and $e$.
 Although $e$ is expected to be a constant
 because the normal velocity of the disk is fixed,
 COR depends on $\gamma$.
 In particular, for small $\cot \gamma$,  
 $e$ decreases as $\cot \gamma$ decreases.
 We will discuss this behavior in the later discussion.

Here, we compare our result with the theory of Maw
 {\it et al.}\cite{stronge,mbf76,mbf81}
 which was consistent
 with experimental data\cite{foerster,lorentz,gorham,mbf76,mbf81}. 
 According to their theory, all the region of the angle of incidence
 can be divided into three regimes.
 For each regime, $\beta$ can be expressed as
\begin{enumerate}
 \item $1/\mu \eta^2 < \cot \gamma$:\hspace{3mm}
\begin{equation}\label{regime1}
\beta=-\cos \omega t_1 - \mu \frac{\beta_x}{\beta_z} e
  \left[1 + \cos \left( \frac{\Omega t_1}{e}+
                   \frac{\pi}{2}(1-e^{-1})\right) \right] \cot \gamma,
\end{equation}
 \item $\beta_x/\beta_z \mu(1+e) < \cot \gamma < 1/\mu \eta^2$:\hspace{3mm}
\begin{equation}\label{regime2}
 \begin{split}
  \beta = &-\cos \omega (t_3-t_2)-\mu \frac{\beta_x}{\beta_z}
  [\cos \omega (t_3-t_2)
  -\cos \Omega t_2 \cos \omega (t_3-t_2)\\
  &+ \frac{\Omega}{\omega} \sin \Omega t_2
  \sin \omega (t_3-t_2)
  + e + \cos \Omega t_3
  ] \cot \gamma,
 \end{split}
\end{equation}
 \item $ \cot \gamma < \beta_x/\beta_z \mu(1+e)$:\hspace{3mm}
\begin{equation}\label{regime3}
\beta=-1 + \mu \frac{\beta_x}{\beta_z}
  \left(1 + e \right) \cot \gamma,
\end{equation}
\end{enumerate}
 where $\mu$ is the coefficient of friction, $\eta$ is the constant
 dependent on Poisson's ratio defined in eq.(\ref{A5}),  
 $\beta_x$ and $\beta_z$ are constants calculated from
 mass, radius, and radii of gyration of material as
 $\beta_x=3.02$ and $\beta_z=1$ as shown in eq.(\ref{appB:betas}).  
 $\Omega$ and $\omega$ are respectively $\pi/2t_c$
 and $(\pi/2\eta t_c)\sqrt{\beta_x/\beta_z}$,
 where $t_c$ is a duration of a collision.
 $t_1$ determined by eq.(\ref{a10})
 is the transition time from initial stick motion to slip motion.
 $t_3$ determined by eqs.(\ref{a21}) and (\ref{a20})
 is the transition time from slip motion to stick motion.
 By calculating $\beta$ at each value of $\cot\gamma$ and interpolating
 them with cubic spline interpolation method,
 we can draw the theoretical curve.

We compare the result of simulation of the oblique impact
 using the random lattice model
 with the theoretical curve(Fig.\ref{theor}).
 Here we used $\eta=1.015$,
 $e=0.8$ which is an average value of COR in Fig.\ref{figure4},
 and $\mu=0.18$ which is decided by comparing the slope
 in the small $\cot \gamma$ region with eq.(\ref{walton}).
 It is found that the result of random lattice model is consistent
 with the theory especially in small $\cot \gamma$ region.
 In the intermediate region,
 the agreement of the data with the theory is worse
 than that of other regions.
 This tendency can be seen in some experimental results
\cite{gorham,mbf81,foerster}. 

 Theoretical result by Maw {\it et al.} suggests that Poisson's ratio
 is not a crucial parameter to determine $\beta$,
 while Poisson's ratio of the triangular lattice is much larger
 than that of the random lattice. To confirm that anomalous behavior
 of the triangular lattice comes from the specific lattice structure
 we simulate the collision by using the square lattice model. 
 By changing the value of spring constants of square lattice disk
 and controlling Poisson's ratio,
 we investigate the dependency of $\beta_0$ on Poisson's ratio. 
 $\beta_0$ are $0.49$ and $0.51$ when $\nu=0.1$ and $\nu=0.3$, respectively.
 From these results, we confirm that Poisson's ratio
 is not a crucial parameter for $\beta_0$. 
\section{Discussions}
Here, we discuss the results of our simulation.
 We change the number of mass points of random lattice model and
 investigate the dependency on the system size.
 As the number of mass points becomes larger, 
 there is a tendency for a graph to be flattened
 in the region of large $\cot \gamma$.
 It can be seen as follows.
 When the model is composed of many mass points,
 irregularity of the surface of the random lattice
 diminishes as the size of the disk increases.
 As a result, $\beta_0$ can take the stable value
 in the large $\cot \gamma$.
 When the number of mass points is larger than $1600$,
 all the mass points in the rectangle cannot be connected by
 the Delaunay triangulation algorithm\cite{delaunay}.
 Hence, our results are restricted to the case
 with the $1600$ mass points as the maximum value.

 We also investigate the influence of roughness of the surface.
 When the standard deviation $\delta$ takes $ 2.15 \times 10^{-2}R$,
 $\beta$ increases monotonously as $\cot \gamma$ increases.
 When $\delta$ takes $ 3.0 \times 10^{-3}R$,
 $\beta$ approaches the stable maximum value $\beta_0 = 0.56$.
 For larger $\delta$, the surface of the rectangle is easy to 
 collapse when the collision occurs.
 As for $\mu_0$,
 $\mu_0$ takes $0.14$ when $\delta$ is $2.15\times 10^{-2}R$
 while $\mu_0$ takes $ 0.18$ when $\delta$ is $3.0\times 10^{-2}R$.
 It can be seen that roughness of the surface make
 the value of $\mu_0$ increase.

 Random lattice model can reproduce experimental tendency 
 in $\beta_0$ and $\mu_0$ with roughness on the surface.
 However,
 the random lattice model cannot reproduce the tendency
 that $\beta$ decreases from the maximum value $\beta_0$
 in the large $\cot \gamma$\cite{labous}. 
 Other mechanisms like sticking or plastic deformation
 on the surface may be important in the large $\gamma$ region. 

 From Fig.\ref{figure3},
 the triangular lattice disk seems to be inadequate
 to reproduce experimental tendency.
 In triangular lattice,
 the shape of the disk is much more like a polygon
 than a circle.
 It may be that polygonal property of the structure causes
 the slip motion of the disk. 

 The decrease of $e$ in the small $\cot\gamma$
 in Fig.\ref{figure4} can be understood as follows.
 In our situation, normal component 
 of initial velocity is fixed to $0.1c$. Thus, the initial kinetic
 energy of the disk becomes larger in the small $\cot\gamma$.
 As a result, the surface of the wall cannot seize the disk
 so that the initial kinetic energy is easy to propagate
 in the horizontal direction to the surface of the wall.

 In the last section we investigated 
 the dependency on Poisson's ratio
 with the aid of the square lattice model.
 In contrast, we change the value of $k_a$ of the triangular lattice
 disk from $1.0 \times m c^2/R^2$ to
 $1.0 \times 10^{2} m c^2/R^2$
 to investigate the dependency on Young's modulus. 
 Although Young's modulus increases by 100-fold,
 the triangular lattice disk remains slippery on the surface.
 From this fact, it can be seen that Young's modulus
 as well as Poisson's ratio are not crucial.
 In addition, we change the value of $k_b$ of triangular lattice disk
 from $1.0 \times 10^{-3} m c^2/R^4$ to $1.0 \times 10^{-1} m c^2/R^4$
 and investigate the effect of the nonlinear term of
 eq.(\ref{interaction}). The change of $k_b$ also does not affect
 the results of triangular lattice disk. 
 It can be seen that the nonlinear term of eq.(\ref{interaction})
 only strengthen the surface of the model and does not
 make the triangular lattice disk rotate after collision.
 In the triangular lattice disk,
 the polygonal property of the surface of triangular lattice
 may affect the results. 

Finally,
 we refer to the connection between $\mu_0$ in eq.(\ref{walton})
 and $\mu$ in eq.(\ref{regime3}).
 In Fig.\ref{theor}, the value of $\mu$ is same as $\mu_0$
 estimated from the slope
 in the range $0 \leq \cot \gamma \leq 2$ in Fig.\ref{figure3}.
 Comparing eq.(\ref{walton}) with eq.(\ref{regime3}),
 we can derive the relation between $\mu_0$ and $\mu$ as
\begin{equation}
\mu \frac{\beta_x}{\beta_z} = \mu_0\left(1+\frac{mR^2}{I}\right).
\end{equation} 
 In the two dimensional binary collision of disks, 
 $1+mR^{2}/I$ can be calculated as $3$ explicitly.
 Meanwhile, $\beta_x/\beta_z$ in our system can be
 calculated as $3.02$.
 Thus, $\mu$ and $\mu_0$ are in our case almost identical.
\section{Conclusion Remarks}
In this paper, 
 we demonstrate the 2-dimensional simulation of the oblique
 impact. Our random lattice model produces
 the same tendency as experimental data qualitatively
 while triangular lattice model can not produce the positive value
 of $\beta_0$. For normal COR, $e$ depends on the initial angle of
 incidence and decreases in the large $\gamma$ when the normal
 component of initial velocity is fixed. For $\beta$,
 we compare our results with Maw's theory of the oblique impact.
 Our result is consistent with their theory
 especially in the large and small region of $\cot \gamma$.

\vspace{5mm}

We appreciate Y. Tanaka, Y. Hayakawa and S. Nagahiro
 for fruitful discussion.
 We also thank M. Doi and S. Takesue for their valuable comments.  
 Parts of numerical computation in this work were carried out
 at the Yukawa Institute Computer Facility.
 This work is partially supported by
 Hosokawa Powder Technology Foundation and Inamori Foundation.


\newpage
\begin{itemize}
\item Fig.1 : The schematic figure of a collision of a disk with a wall.
\item Fig.2 : The elastic disk and wall consisted of random lattice system.
\item Fig.3 : Interaction between surface particles of the disk and the wall.
\item Fig.4 : The bands of random lattice (a) before and (b) after stretch.
\item Fig.5 :  The schematic figures of (a) triangular lattice disk and (b) square lattice disk.
\item Fig.6 : The relation between cotangent of angle of incidence
   $\gamma$ and $\beta$. Cross points are the results of the random
 lattice disk. Stars are the result when the random disk
 has no roughness. Plus points are the results of the triangular lattice
 disk. Dashed and dot-dash lines are eq.(\ref{walton}).
\item Fig.7 : The relation between cotangent of angle of incidence
   $\gamma$ and COR $e$.
\item Fig.8 : The relation between $\cot\gamma$ and $\beta$.
 Cross points are the numerical results of the random lattice model.
 Solid line is the theoretical curve.
\item Fig.9 : The schematic figure of the disk and the wall. A cross
 in a circle represents a center of mass of each body.
\end{itemize}
\newpage
\appendix
\section{The Theoretical Description of the Oblique Impact}\label{appendixA}
Here, we review and rewrite a theory of
 the oblique impact\cite{stronge,mbf76,mbf81} for our investigation.
 Let a disk with the radius $R$ and a rectangle
 with the hight $2R$ and the width $8R$
 be in contact each other as depicted in Fig.(\ref{appB:app_fig1}).
 They have masses $M$ and $M^{'}$
 and their radii of gyration
 $\hat k_r = R/\sqrt{2}$ and
 $\hat k^{'}_r = R \sqrt{17/3}$
 around their centers of mass, respectively.
 The prime denotes parameters for the rectangle.
 The position of the contact point $C$
 is denoted as $(r_x,r_z)=(0,-R)$
 or $(r^{'}_x,r^{'}_z)=(0,R)$
 which are measured from the centers of mass
 of each colliding body.
 Here, $v_{i}$ and $u_{i}$ $(i=x,z)$
 are relative velocity and relative displacement
 at the contact point, respectively.
 We assume that both normal and tangential elements of
 the compliance are proportional to the compression.
 We also introduce the normal stiffness during compression $\kappa$,
 and the tangential stiffness during compression $\kappa / \eta^{2}$
 for the disk. 
 The equation of motion of the displacements, thus,  becomes
\begin{equation}
 \left(
 \begin{array}{c}
  \Ddot u_x\\
  \Ddot u_z
 \end{array}
 \right)=
 -m^{-1} \kappa 
  \left[
 \begin{array}{cc}
  \beta_x \eta^{-2} & 0 \\
  0 & \beta_z 
 \end{array}
  \right]
  \left(
 \begin{array}{c}
  u_x\\
  u_z
 \end{array}
  \right),\label{appB:eq1}
\end{equation}
 where $\Ddot u_x=d^{2}u_x/dt^{2}$, and
\begin{equation}\label{appB:betas}
\beta_x = 1+\frac{m r_z^{2}}{M {\hat k}^2_r}
+\frac{m {r_z^{'}}^{2}}{M^{'} \hat {k^{'}}^2_r},
\hspace{5mm}
\beta_z = 1+\frac{m r_x^{2}}{M {\hat k}^2_r}
+\frac{m {r_x^{'}}^{2}}{M^{'} \hat {k^{'}}^2_r},
\end{equation}
 where $1/m=1/M + 1/M^{'}$.
 In our situation, $\beta_x$ and $\beta_z$ can be calculated
 as $\beta_x \simeq 3.02$ and $\beta_z = 1$.

Equation of motion (\ref{appB:eq1}) has two characteristic frequencies,
 $\Omega$ and $\omega$, which are expressed as
\begin{equation}
\Omega \equiv \sqrt{\frac{\beta_{z} \kappa}{m}},\hspace{5mm}
 \omega \equiv \sqrt{\frac{\beta_{x} \kappa}{\eta^{2} m}}
=\frac{1}{\eta} \sqrt{\frac{\beta_x}{\beta_z}} \Omega\label{appB:Omg}
\end{equation}
in the normal and tangential direction
 to the wall, respectively.
 One can also express them using $t_{c}$, which is
 the moment when the normal velocity of compression becomes $0$
($\Dot{u_{z}}(t_{c})=0$), as $\Omega=\pi/2 t_c$
 and $\omega= \left(\pi/2 \eta t_c \right) \sqrt{\beta_x/\beta_z}$.
According to Johnson\cite{johnson}, stiffnesses in the normal and
 tangential direction, $\kappa$ and $\kappa / \eta^{2}$, can be expressed  
 using Young's modulus $E$, the radius of punch $a$, and Poisson's ratio
 $\nu$ as 
\begin{equation}
\kappa = \frac{E a}{1 - \nu^{2}},\hspace{5mm}
 \kappa/\eta^2 = \frac{2 E a}
{\left(2-\nu\right)\left(1+\nu\right)}\label{appB:kappas}.
\end{equation}
 Thus, $\eta$ can be expressed only by the Poisson's ratio,
\begin{equation}
 \eta = \sqrt{\frac{2-\nu}{2 (1-\nu)}}\label{A5}.
\end{equation}

Here we define the coefficient of restitution
 for subsequent discussion.
 We assume collision starts at $t=0$
 and compression and restitution periods terminate at $t=t_c$
 and $t=t_f$, respectively.
 The coefficient of restitution is defined as
\begin{equation}
 e_*=\frac{p_z(t_f)-p_z(t_c)}{p_z(t_c)}=-\frac{v_z(t_f)}{v_0},
\end{equation}
 where $p_z(t)$ is the normal impulse, $v_z(t)$ is the normal velocity,
 and $v_0$ is the initial normal velocity. This leads to
 $p_z(t_f) = (1+e_*) p_z(t_c)$ and $t_f=(1+e_*)t_c$.
In the later discussion,
 we assume that the effect of the coefficient of restitution $e_*$
 is obtained by changing the stiffness of the normal compliant element
 from $\kappa$ to $\kappa/e_*^{2}$ at $t=t_c$.
 Thus, from eq.(\ref{appB:Omg}),
 the normal frequency increases from $\Omega$ to $\Omega/e_*$ at $t=t_c$.
\subsection{Normal components of velocity and Force}
Now, we solve eq.(\ref{appB:eq1}) and obtain
 normal and  tangential components of velocity and force.
 Assuming $v_{z}(t)+\Dot u_z(z)=0$ during contact,
 we obtain
\begin{equation}\label{appB:eq7}
v_{z}(t) =
\left\{
 \begin{array}{ll}
 v_z(0) \cos \Omega t & 0 \leq t \leq t_c\\
 e_{*}v_z(0) \cos( \frac{\Omega t}{e_{*}}+\frac{\pi}{2}(1-e^{-1}_{*}) )
 & t_c \leq t \leq t_f.
 \end{array}
\right.
\end{equation}
Assuming that the normal frequencies
 during restitution is $\Omega/e_{*}$
 and the initial conditions $v_z(t_c)=0$ and $v_z(t_f)=-e_{*}$,
 we reach the exact form of solution during restitution period. 
 (\ref{appB:eq7}) is continuous at $t=t_{c}$.
 By differentiating these expressions,
 we also obtain the displacement $u_z(t)$ , the force $F_z(t)$,
 and the impulse $p_z(t)$ as described
 in Table \ref{appB:appendixtable1}.
\subsection{Tangential Components of Velocity and Force}
 We assume that a disk sticks or slips
 on the surface of a rectangle during collision
 and starts sticking at $t=t_2$.
 At $t=t_2$, the relation between $F_x$ and $F_z$
 becomes $|F_x| < \mu F_z$.
 We calculate the expressions for tangential components
 for sticking and slipping separately. 

 While a disk slips on the surface of a rectangle, 
 the relation between $F_x$ and $F_z$ is $|F_x|/F_z = \mu$
 and the tangential velocity
 is changed by a impulse arising from contact force.
 Thus, The change of velocity can be described as
\begin{equation}
 \left(
 \begin{array}{c}
  dv_x/dp_3\\
  dv_z/dp_3
 \end{array}
 \right)=
 m^{-1}
  \left[
 \begin{array}{cc}
  \beta_x \eta^{-2} & 0 \\
  0 & \beta_z 
 \end{array}
  \right]
  \left(
 \begin{array}{c}
  -\mu sgn(v_x+\dot u_x)\\
  1
 \end{array}
  \right)
\hspace*{10mm}
t < t_2,\label{appB:velchg}
\end{equation}
 where $sgn(x)=+1$ for $x>0$ and $sgn(x)=-1$ for $x<0$. 

 When a disk starts sticking at $t=t_2$,
 the tangential oscillation starts with frequency $\omega$.
 We assume $v_x(t)+\Dot u_x(t)=0$.
 By solving the equation of $u_x(t)$,
 we can obtain the tangential components
 of displacement, velocity, and contact force as
\begin{eqnarray}
u_x(t)&=& u_x(t_2) \cos \omega(t-t_2)
-\omega^{-1} v_x(t_2) \sin \omega(t-t_2)\notag\\
v_x(t)&=&\omega u_x(t_2) \sin \omega(t-t_2)
+v_x(t_2) \cos \omega(t-t_2)\label{appB:f1}\\
F_x(t)&=&m \beta^{-1}_x \omega^{2} u_x(t_2)\cos \omega(t-t_2)
-m \beta^{-1}_x \omega v_x(t_2)\sin \omega(t-t_2)
\hspace*{10mm}
 t \geq t_2\notag.
\end{eqnarray}
%
\subsection{Obtaining the transition time, $t_1$}
 We think the situation that stick begins
 at the initial instant contact, {\it i.e.} $t_2=0$,
 and slip begins at $t=t_1$.
 From (\ref{appB:f1}) with the condition $u_x(0)=0$
 and $F_z(t)$ described in the Table \ref{appB:appendixtable1},
 $t_1$ can be obtained solving the equation
\begin{equation}
\frac{|F_x(t_1)|}{\mu F_z(t_1)}
= \begin{cases}
   \displaystyle \frac{1}{\eta^2} \frac{v_x(0)}{\mu v_z(0)}
   \frac{\Omega}{\omega} \frac{\sin \omega t_1}{\sin \Omega t_1}=1
   &0 \leq t_1 < t_c\\
   \displaystyle \frac{1}{\eta^2} \frac{v_x(0)}{\mu v_z(0)}
   \frac{\Omega}{\omega}
   \frac{\sin \omega t_1}{\sin (\frac{\Omega t_1}{e_*}+
   \frac{\pi}{2}(1-e^{-1}_{*}))}=1
   &t_c \leq t_1 < t_f\label{a10}.
   \end{cases}
\end{equation}
 Solving these equation numerically, we obtain $t_1$.
It should be noted that there are two conditions
 if $t_1$ is greater or smaller than $t_c$.

The process of initial stick takes place if $t_1>0$, {\it i.e.} if in the limit as
$t_1\rightarrow 0$ the force ratio
 between the tangential and normal component is smaller than $\mu$.
 This requires 
\begin{equation}
\frac{v_x(0)}{v_z(0)} < \mu \eta^2.
\end{equation}
\subsection{Three Regimes of The Angle of Incidence}
 Here, we divide all region of the angle of incidence into three
 regimes and calculate the tangential component of terminal velocity
 of collision for each regime.   
\vspace*{3mm}
 
(i)\hspace*{3mm}{\bf small angle of incidence:} $v_x(0)/v_z(0) < \mu \eta^2$

In this regime, initial stick continues until $t=t_1$
 and slip terminates at $t=t_f$.
 At time $t_1$, the
tangential component of the relative velocity
 is $v_x(t_1)=v_x(0) \cos\omega t_1$.
From eq.(\ref{appB:velchg}),
 the terminal tangential velocity can be expressed as
\begin{equation}
v_x(t_f)=v_x(0)\cos\omega t_1
-\frac{\mu  \beta_x}{m}[p_z(t_f)-p_z(t_1)],\label{appB:reg1}
\end{equation}
 where $p_z(t_f)-p_z(t_1)$ can be expressed as
\begin{equation}
p_z(t_f)-p_z(t_1)=-m \frac{v_z(0)}{\beta_z}
e_{*}\{1+\cos(\frac{\Omega t_1}{e_*}+\frac{\pi}{2}(1-e^{-1}_{*}))\} 
\end{equation}
 from Table \ref{appB:appendixtable1}.
 Dividing eq.(\ref{appB:reg1}) by $-v_x(0)$ leads to
\begin{equation}
\beta=-\cos \omega t_1(\gamma) - \mu \frac{\beta_x}{\beta_z} e
  \left[1 + \cos \left( \frac{\Omega t_1(\gamma)}{e}+
                   \frac{\pi}{2}(1-e^{-1})\right) \right]
 \cot \gamma,
\end{equation}
 where $\beta$ is $-v_x(t_f)/v_x(0)$ and $\cot \gamma$ is 
 $v_z(0)/v_x(0)$.
\vspace*{3mm}

(ii)\hspace*{3mm}{\bf intermediate angle of incidence:}
 $\mu \eta^2 < v_x(0)/v_z(0) < \mu(1+e_*)\beta_x/\beta_z$

In this regime, the disk initially slips and begins to stick
 at $t=t_2$. After the period of sticking, the disk begins
 to slip again at $t=t_3$.
 In the period $t<t_2$,
 the tangential component of relative velocity is written as
\begin{equation}
v_x(t)=v_x(0)-\frac{\mu \beta_x}{m} p_z(t).\label{appB:2reg}
\end{equation}

 Here let us calculate $t_2$ and $t_3$. 
 At $t=t_2$,
 subsequent sliding and stick give the same rate of change for the
 tangential force:
\begin{equation}
\displaystyle \lim_{\epsilon \to 0} \left|\frac{dF_x(t_2+\epsilon)}{d
 \epsilon}\right|
=\lim_{\epsilon \to 0}\mu \frac{dF_z(t_2+\epsilon)}{d \epsilon}.\label{appB:2reg2}
\end{equation}
This is the condition which determines $t_2$. To simplify this
condition, we need to obtain the exact forms of
 $dF_x(t)/dt$ and  $dF_z(t)/dt$.
 
For tangential components of force, if one differentiate
eq.(\ref{appB:f1}) by $t$, we obtain
\begin{equation}
\frac{dF_x(t)}{dt}=\frac{m \omega^3 u_x(t_2)}{\beta_x} \sin \omega(t-t_2)
-\frac{m \omega^2 v_x(t_2)}{\beta_x} \cos \omega(t-t_2)\hspace*{10mm}
t>t_2.\label{appB:f_xdt}
\end{equation}
Here,
 $v_x(t)$ is represented by eq.(\ref{appB:2reg})
 and with the aid of Table \ref{appB:appendixtable1},
 and $u_x(t)$ is obtained from $v_x(t)+\dot u_x(t)=0$.
 Thus, the explicit expressions are
\begin{eqnarray}
v_x(t_2)
&=& \begin{cases}
   v_x(0)-\mu \frac{\beta_x}{\beta_z} v_z(0)[1-\cos \Omega t_2]
   & t_2 \le t_c\\
   v_x(0)-\mu \frac{\beta_x}{\beta_z} v_z(0)
   \left[1-\cos\left(\frac{\Omega t_2}{e_*}
   +\frac{\pi}{2}(1-e_*^{-1})\right)\right]
   &t_2 > t_c
   \end{cases}\label{appB:v_xt2}\\
u_x(t_2)
&=& \begin{cases}
   \mu \frac{\beta_x \Omega v_z(0)}{\beta_z \omega^2} \sin \Omega t_2,
   & t_2 \le t_c\\
   \mu \frac{\beta_x \Omega v_z(0)}{\beta_z \omega^2}
   \sin\left(\frac{\Omega t_2}{e_*}+\frac{\pi}{2}(1-e_*^{-1})\right)
   & t_2 > t_c\notag
   \end{cases}
\end{eqnarray}

For normal components of force, by differentiating the expressions
 of normal components in Table \ref{appB:appendixtable1}, we can obtain 
\begin{equation}
\frac{dF_z(t_2)}{dt}
 =\begin{cases}
    -\beta^{-1}_z \Omega^{2} m v_z(0) \cos \Omega t_2
    & t_2 \leq t_c\\
    -\frac{\Omega^2 m v_z(0)}{\beta_z e_*}
    \cos(\frac{\Omega t_2}{e_*}+\frac{\pi}{2}(1-e^{-1}_*))
    & t_2 \geq t_c\label{appB:f_zdt}
  \end{cases}
\end{equation}

From (\ref{appB:f_xdt}),(\ref{appB:v_xt2}), and (\ref{appB:f_zdt}),
 eq.(\ref{appB:2reg2}) leads to
\begin{eqnarray}
\Omega t_2 &=& \arccos \left(
 \frac{v_x(0)/\mu v_z(0)-\beta_x/\beta_z}{\eta^2-\beta_x/\beta_z}
\right)\hspace*{10mm}\frac{v_x(0)}{v_z(0)} \leq \mu \frac{\beta_x}{\beta_z}
\label{a20}\\
\frac{\Omega t_2}{e_*} &=& -\frac{\pi}{2}(1-e^{-1}_*)
+\arccos \left(
 \frac{v_x(0)/\mu v_z(0)-\beta_x/\beta_z}{\eta^2 e^{-1}_* - e_* \beta_x/\beta_z}
\right)\hspace*{10mm}\frac{v_x(0)}{v_z(0)} > \mu \frac{\beta_x}{\beta_z}\notag
\end{eqnarray}

In the period $t_2 < t < t_3$, the velocity and the force
 are expressed as eq.(\ref{appB:f1}).
 This period of stick terminates and
 slip begins at time $t=t_3$. $t_3$ can be determined
 by the condition $|F_x|/F_z=\mu$.
 From eq.(\ref{appB:f1}) and Table \ref{appB:appendixtable1},
 this condition leads to
\begin{equation}
\left| \frac{\Omega u_x(t_2)}{\mu v_z(0)}\cos \omega(t_3-t_2)
-\frac{\Omega v_x(t_2)}{\omega \mu v_z(0)}\sin \omega(t_3-t_2) \right|
=\eta^2 \sin\left[\frac{\Omega t_3}{e_*}+\frac{\pi}{2}(1-e^{-1}_*)\right]
\label{a21}.
\end{equation}
Solving this equation numerically, we obtain $t_3$.
The final tangential velocity is expressed as
\begin{equation}
v_x(t_f)=v_x(t_3)-\mu \beta_x m^{-1}[p_z(t_f)-p_z(t_3)]
\label{appB:v_xt_f},
\end{equation}
 where $p_z(t)$ is expressed in Table \ref{appB:appendixtable1}.
Dividing (\ref{appB:v_xt_f}) by $-v_x(0)$, we obtain
\begin{equation}
 \begin{split}
  \beta = &-\cos \omega (t_3-t_2)-\mu \frac{\beta_x}{\beta_z}
  [\cos \omega (t_3 - t_2)
  -\cos \Omega t_2 \cos \omega (t_3 - t_2)\\
  &+ \frac{\Omega}{\omega} \sin \Omega t_2
  \sin \omega (t_3 - t_2)
  + e + \cos \Omega t_3  ] \cot \gamma.
 \end{split}
\end{equation}
\vspace*{3mm}

(iii)\hspace*{3mm}{\bf Large angle of incidence:}
 $v_x(0)/v_z(0) > \mu(1+e_*)\beta_x/\beta_z$

In this regime, slip does not cease before separation: $t_2 > t_f$.
 At separation, the tangential velocity $v_x(t_f)$ if as follows:
\begin{equation}
v_x(t_f)=v_x(0)+\mu \beta_x m^{-1}(1+e_*) p_z(t_c)\label{appB:reg3},
\end{equation}
 where $p_z(t_c)=-m v_z(0)/\beta_z$.
 Thus, dividing (\ref{appB:reg3}) by $-v_x(0)$, we obtain
\begin{equation}
\beta=-1 + \mu \frac{\beta_x}{\beta_z}
  \left(1 + e \right) \cot \gamma.
\end{equation}
\section{Poisson's Ratio of The Square Lattice System}\label{appendixB}
In this Appendix,
 we derive the relation between elastic constants in continuum limit
 and spring constants of a two-dimensional square lattice
 with nearest neighbor coupling $k_1$ and next nearest neighbor
 coupling $k_2$.
 The elastic tensor $C_{ijkl}$ for the two dimensional square lattice
 is represented as

\begin{eqnarray}
C_{xxxx}=C_{yyyy}&=&k_{1}+k_{2} \label{eq1}, \\ 
C_{xxyy}=C_{yyxx}&=&C_{xyyx}=C_{yxxy}=C_{xyxy}=C_{yxyx}=k_{2}\label{eq1-1},
\end{eqnarray}
and the other coefficients are zero\cite{feynman}.

Using the elastic tensor $C_{ijkl}$
 and the strain tensor $u_{ij}$ and $u_{kl}$,
 The free energy of the system $U$ is represented as
\begin{equation}
U=\frac{1}{2} C_{ijkl} u_{ij} u_{kl}.
\end{equation}
Thus, we obtain the stress tensor $\sigma_{ij}$ as
\begin{equation}
\sigma_{ij} = \frac{\partial U}{\partial u_{ij}} = C_{ijkl} u_{kl}.
\end{equation}

Now we introduce the unit vector $\mathbf{n}$
 in the axial direction of the rod.
 When we pull both sides of the rod with the pressure $p$,
 the relation 
\begin{equation}\label{eq2}
\sigma_{ik} = p n_{i} n_{k}
\end{equation}
 holds.

From Eqs (\ref{eq1}), (\ref{eq1-1}) and (\ref{eq2}), 
the explicit expressions of the stress tensors become
\begin{eqnarray}
\sigma_{xx} &=& C_{xxxx} u_{xx} + C_{xxyy} u_{yy}
            = (k_{1}+k_{2}) u_{xx} + k_{2} u_{yy} = p n_{x}^2 \\
\sigma_{yy} &=& C_{yyxx} u_{xx} + C_{yyyy} u_{yy}
            = k_{2} u_{xx} + (k_{1}+k_{2}) u_{yy} = p n_{y}^2 \\
\sigma_{xy} &=& C_{xyxy} u_{xy} + C_{xyyx} u_{yx}
            = 2 k_{2} u_{xy} = p n_{x} n_{y}.
\end{eqnarray}
From these equations, we obtain the expressions of the strain tensors,
\begin{eqnarray}
u_{xx} &=& p \frac{(k_{1} + k_{2}) n_{x}^2 - k_{2} n_{y}^2}
{k_{1} (k_{1} + 2 k_{2})} \\
u_{yy} &=& p \frac{(k_{1} + k_{2}) n_{y}^2 - k_{2} n_{x}^2}
{k_{1} (k_{1} + 2 k_{2})} \\
u_{xx} &=& p \frac{n_{x} n_{y}}{2 k_{2}}.
\end{eqnarray}

The strain in the direction of $\mathbf{n}$ is expressed
 as $u=u_{ik} n_{i} n_{k}$.
 Thus, we have
\begin{eqnarray}
u &=& u_{xx} n_{x}^{2} + 2 u_{xy} n_{x} n_{y} + u_{yy} n_{y}^{2} \\
&=& \left\{ \frac{k_{1}+k_{2}}{k_{1}(k_{1} + 2 k_{2})}
+\frac{k_{1}-2k_{2}}{k_{1}k_{2}}n_{x}^{2} n_{y}^{2} \right\}p.
\end{eqnarray}

Therefore we obtain Young's modulus as
\begin{equation}
\frac{1}{E}=\frac{k_{1}+k_{2}}{k_{1}(k_{1} + 2 k_{2})}
+\frac{k_{1}-2k_{2}}{k_{1}k_{2}}n_{x}^{2} n_{y}^{2}
\end{equation}

On the other hand, the Poisson's ratio $\nu$ is defined as the ratio
 of the normal strain to the vertical strain.
 The latter is described as
\begin{eqnarray}
u_{t} &=& u_{xx} n_{y}^{2} - 2 u_{xy}n_{x}n_{y}+u_{yy}n_{x}^{2}\\
      &=& \left\{\frac{2(k_{1}+2k_{2})n_{x}^2 n_{y}^2-k_{2}}{k_{1}(k_{1}+2k_{2})}-\frac{n_{x}^2 n_{y}^2}{k_{2}} \right\} p.
\end{eqnarray}
 
Thus, Poisson's ratio is given by 
\begin{eqnarray}
\nu = \frac{k_{2}^2 + (k_{1}^2-4k_{2}^2)n_{x}^2 n_{y}^2}
{k_{2}(k_{1}+k_{2})+(k_{1}^2-4 k_{2}^2) n_{x}^2 n_{y}^2}.
\end{eqnarray}
\newpage
\newpage
\begin{figure}[thbp]
  \begin{center}
   \includegraphics[width=0.7\textwidth,angle=270]{random.ps}
   \caption{H. Kuninaka and H. Hayakawa}
   \label{figure1}
  \end{center}
\end{figure}
\newpage
\begin{figure}[h]
  \begin{center}
   \includegraphics[width=1.0\textwidth]{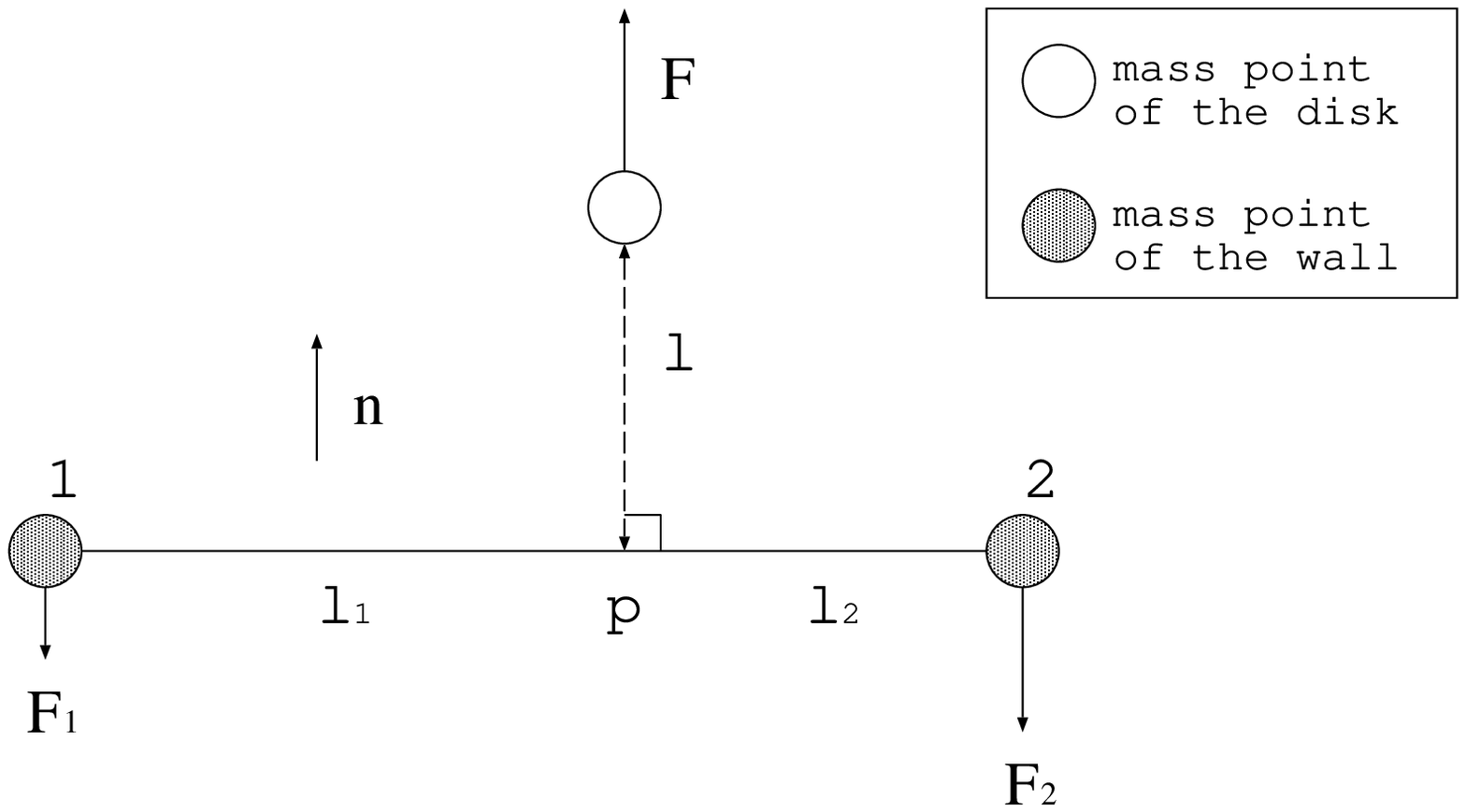}
   \caption{H. Kuninaka and H. Hayakawa}
   \label{figure2}
  \end{center}
\end{figure}
\newpage
\begin{figure}[h]
  \begin{center}
   \includegraphics[width=1.0\textwidth]{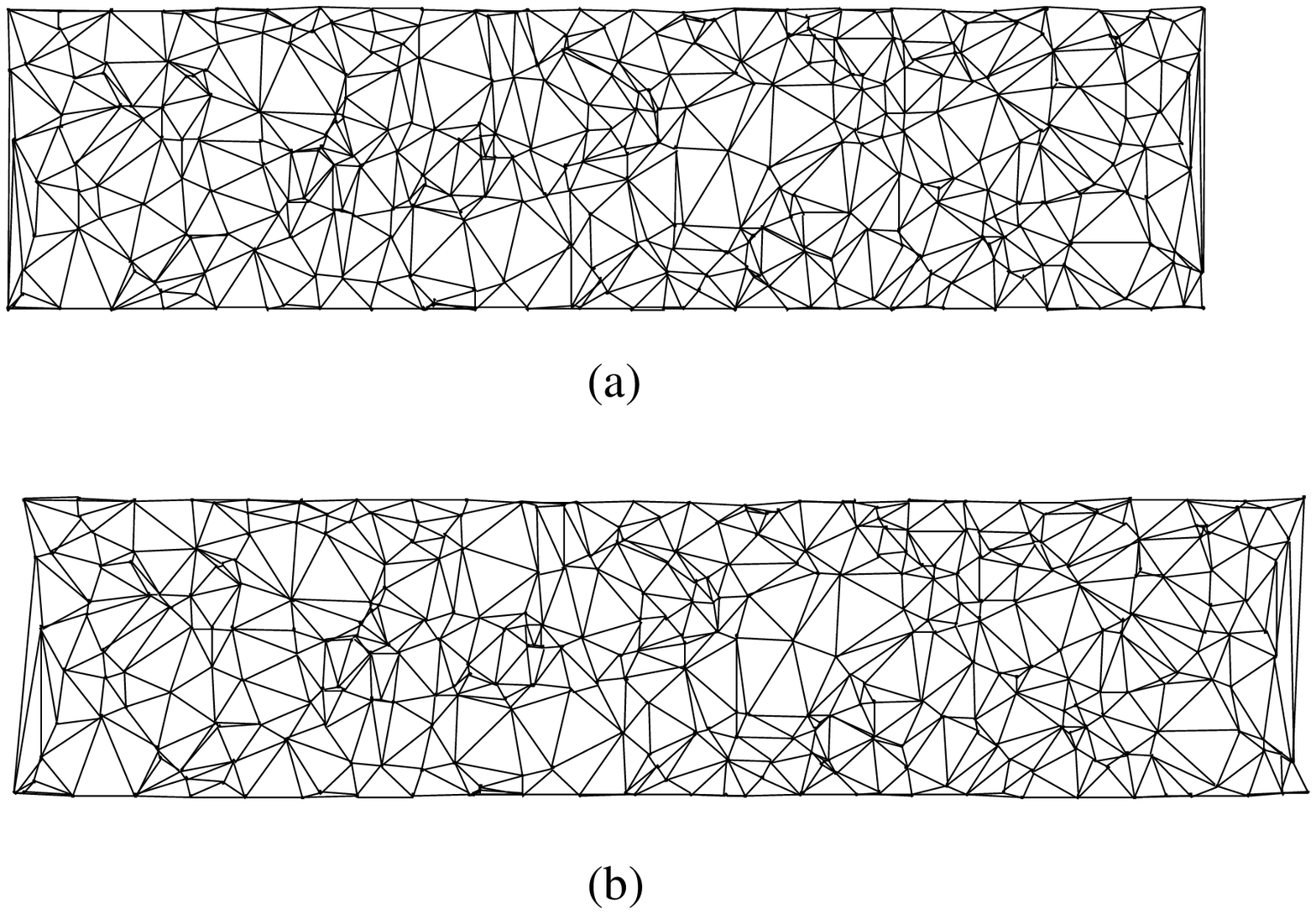}
   \caption{H. Kuninaka and H. Hayakawa}
   \label{ranst}
  \end{center}
\end{figure}
\newpage
\begin{figure}[hbtp]
  \begin{center}
   \includegraphics[width=1.0\textwidth]{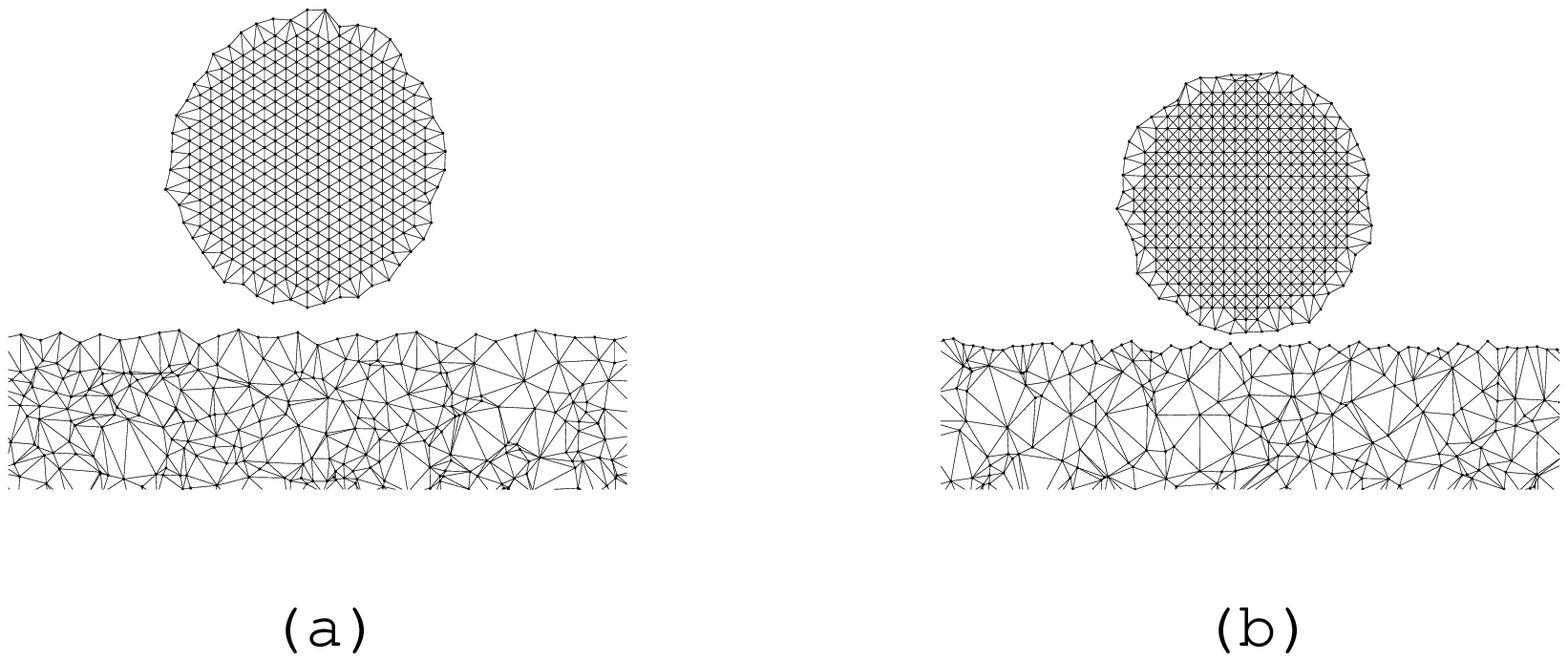}
   \caption{H. Kuninaka and H. Hayakawa}
   \label{lattice}
  \end{center}
\end{figure}
\newpage
\begin{figure}[hp]
  \begin{center}
   \includegraphics[width=1.0\textwidth]{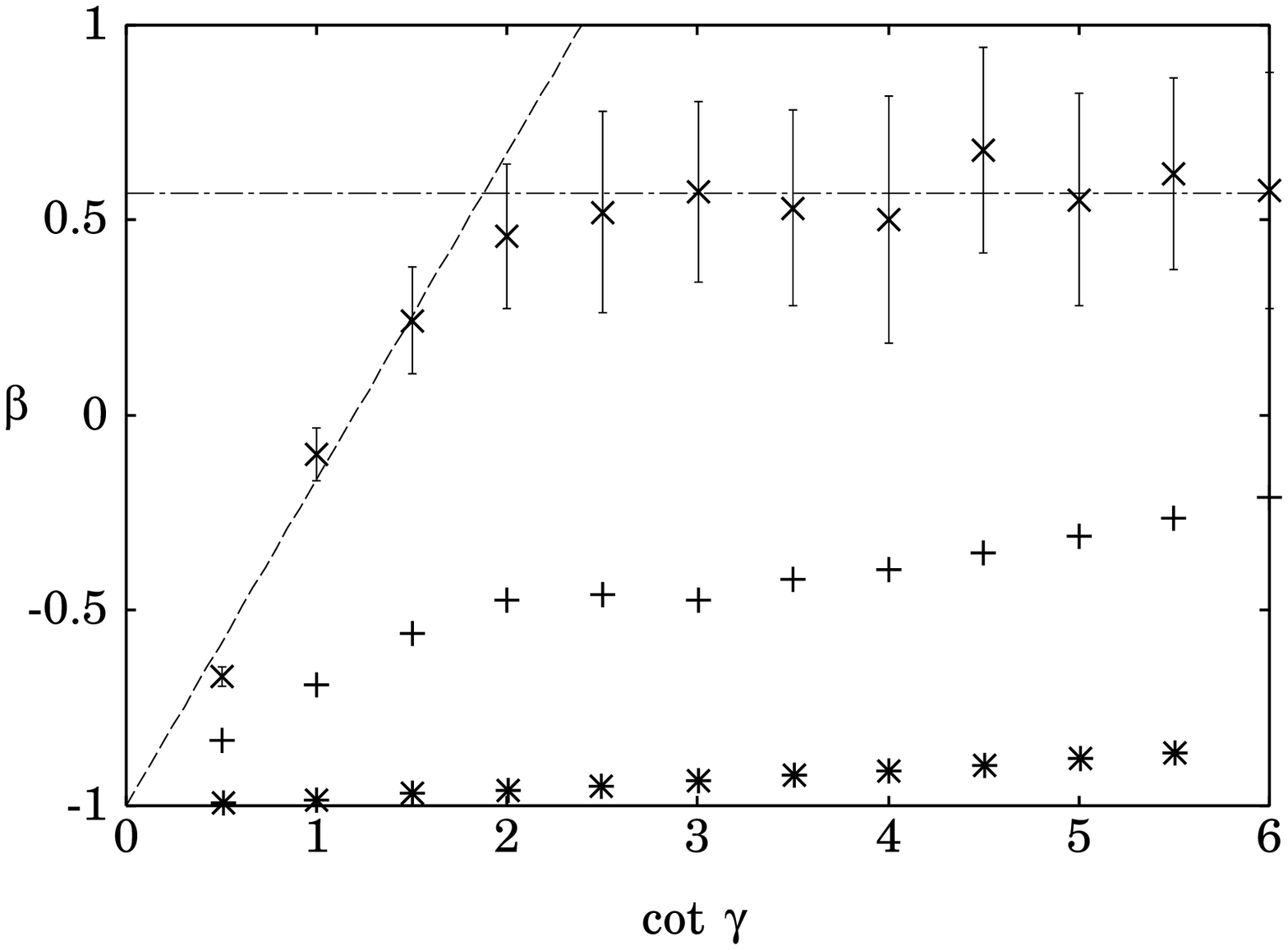}
   \caption{H. Kuninaka and H. Hayakawa}
   \label{figure3}
  \end{center}
\end{figure}
\newpage 
\begin{figure}[hp]
  \begin{center}
   \includegraphics[width=1.0\textwidth]{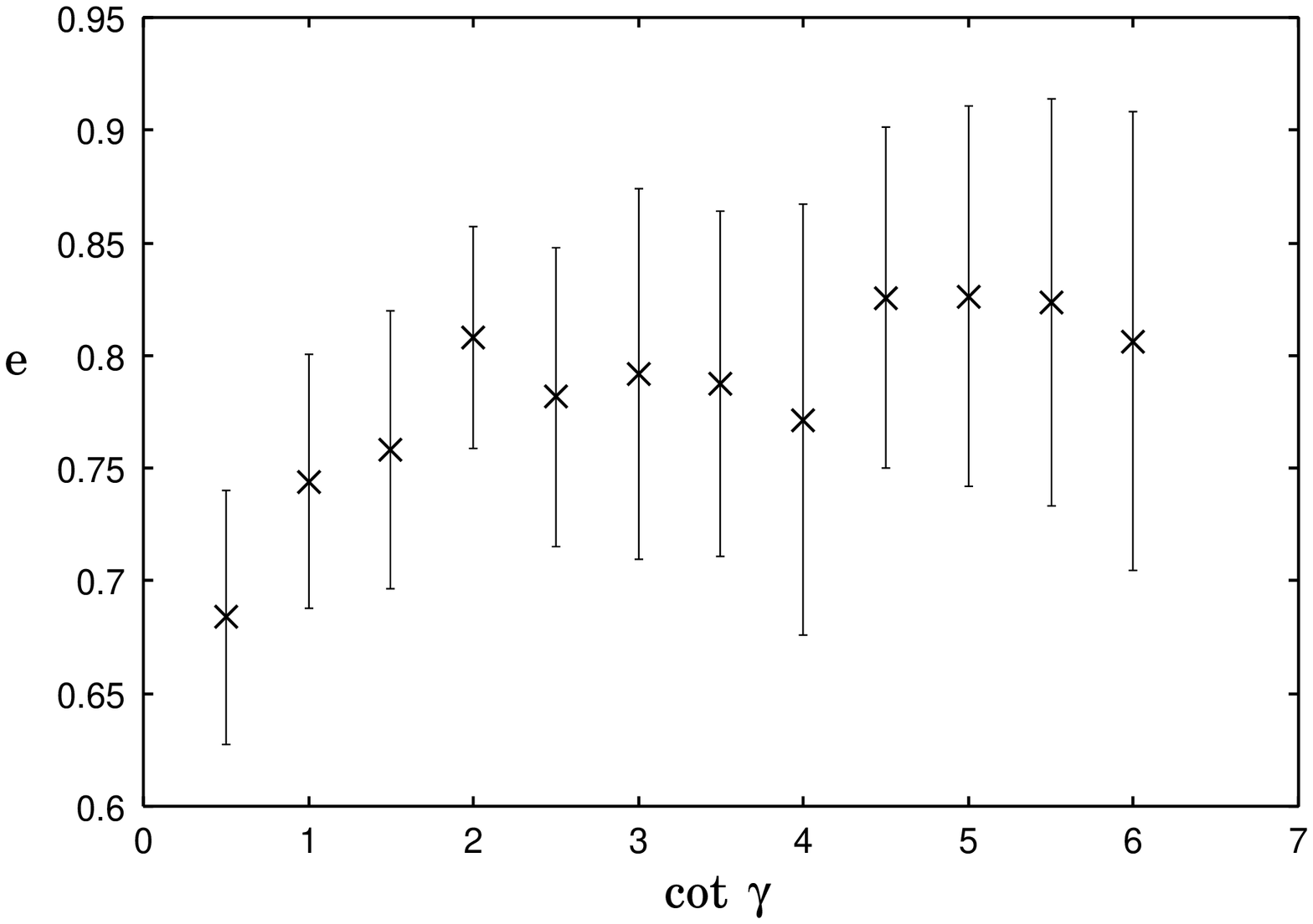}
   \caption{H. Kuninaka and H. Hayakawa}
   \label{figure4}
  \end{center}
\end{figure}
\newpage
\begin{figure}[h]
  \begin{center}
   \includegraphics[width=1.0\textwidth]{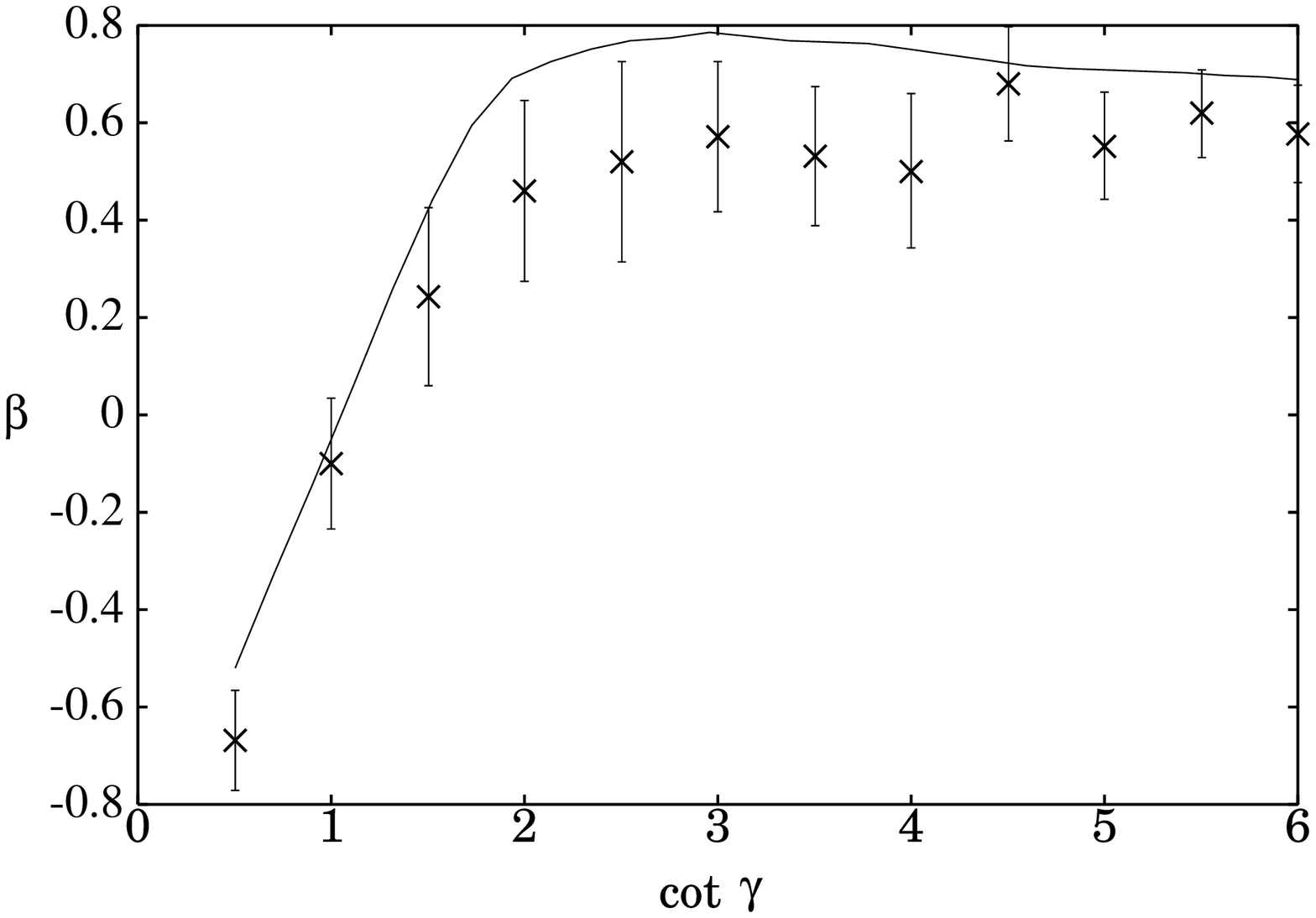}
   \caption{H. Kuninaka and H. Hayakawa}
   \label{theor}
  \end{center}
\end{figure}
\newpage
\begin{figure}[hbtp]
\begin{center}
\includegraphics[width=1.0\textwidth]{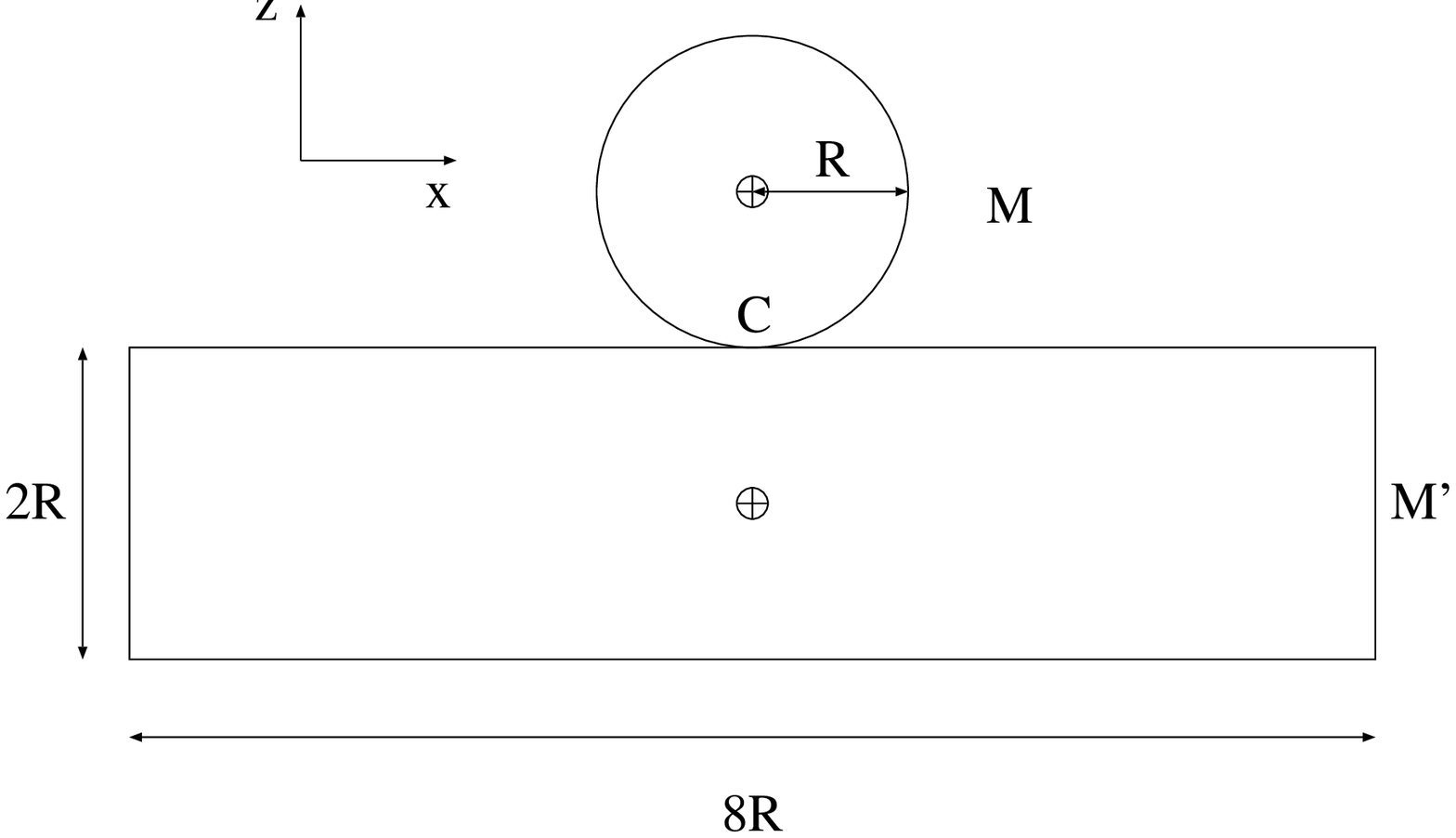}
\end{center}
\caption[]{H.Kuninaka and H. Hayakawa}
\label{appB:app_fig1}
\end{figure}
\newpage
\begin{table}[p]
\caption{Normal displacement, velocity, force, and impulse}
\begin{center}
\begin{tabular}{lll}\hline
Quantity & Compression $(0 \le t \le t_c)$& Restitution $(t_c \le t \le t_f)$\\
\hline
Displacement
&$u_z(t)=-\Omega^{-1} v_z(0) \sin \Omega t$
&$u_z(t)=-e^{2}_* \Omega^{-1} v_z(0) \sin \left(\frac{\Omega t}{e_*}
 + \frac{\pi}{2}\left(1-e_*^{-1}\right)\right)$\\ 
Velocity
&$v_z(t)=v_z(0) \cos \Omega t$
&$v_z(t)=e^{2}_* v_z(0) \cos \left(\frac{\Omega t}{e_*}
 + \frac{\pi}{2}\left(1-e_*^{-1}\right)\right)$\\ 
Force
&$F_z(t)=-\frac{m \Omega v_z(0)}{\beta_z} \sin \Omega t \ge 0$
&$F_z(t)=-\frac{m \Omega v_z(0)}{\beta_z} \sin \left(
\frac{\Omega t}{e_*}
 + \frac{\pi}{2}\left(1-e_*^{-1}\right)\right) \ge 0$\\
Impulse
&$p_z(t)=-\frac{m v_z(0)}{\beta_z} \left(1-\cos \Omega t\right)$
&$p_z(t)=-\frac{m v_z(0)}{\beta_z} \left(1-e_* \cos \left(
\frac{\Omega t}{e_*}
 + \frac{\pi}{2}\left(1-e_*^{-1}\right)\right)\right)$\\ 
\hline
\end{tabular}
\end{center}
\label{appB:appendixtable1}
\end{table}
\newpage
\end{document}